  \documentclass[prl,aps,twocolumn,showpacs]{revtex4}
\usepackage[dvips]{graphicx}
\usepackage{dcolumn}%
\usepackage{amsmath}%
\usepackage{amsfonts}%
\usepackage{amssymb}
\providecommand{\U}[1]{\protect\rule{.1in}{.1in}}
\newcommand{\be}{\begin{equation}}
\newcommand{\ee}{\end{equation}}
\newcommand{\bea}{\begin{eqnarray}}
\newcommand{\eea}{\end{eqnarray}}
\newcommand{\nn}{ \nonumber}
\newcommand{\ds}{\displaystyle}
\begin{document}

\title{Quantum oscillations in the high frequency magnetoacoustic response\\ of a quasi-two-dimensional metal }

\author{ Natalya A. Zimbovskaya$^{1}$ and Godfrey Gumbs$^2$}

\affiliation{$^1$Department of Physics and  Electronics, University of Puerto Rico, 100 CUH Station, Humacao, PR 00791, \\ 
Institute for Functional Nanomaterials, University of Puerto Rico, San Juan, PR 00931, }
\affiliation{ 
 $^2$Department of Physics and Astronomy, Hunter College of the City University \\ of New York,  695 Park Avenue, New York, NY 10065 }

\begin{abstract} 
 In this work we present the results of theoretical analysis of magnetic quantum oscillations of the velocity and attenuation of high frequency ultrasound waves traveling in quasi-two-dimensional conductors. We chose a geometry where the  wave vector of the longitudinal sound wave and the external magnetic field are directed along the axis of symmetry of the Fermi surface. Assuming a moderately weak Fermi surface corrugation we showed that the oscillating correction to the sound velocity may include a special term  besides an ordinary contribution originating from quantum oscillations of the charge carriers density of states at the Fermi surface. This additional term is generated by a ``phase stability" resonance occurring when the charge carriers velocity in the direction of the wave propagation equals the sound velocity. The two oscillating contributions to the sound velocity are shown to differ in phase and shape, and they may have the same order of magnitude. The appearance of the extra term may bring significant changes in magnetic quantum oscillations of the velocity of sound in Q2D conductors, especially at low temperatures.  
 \end{abstract}


\date{\today}
\maketitle

 
 It is a common knowledge that quantization of conduction electrons motion in strong magnetic fields gives rise to oscillations of the electron density of states (DOS) at the the Fermi surface (FS) of a metal. These quantum oscillations generate  several effects such as de Haas-van Alphen oscillations in the magnetization and Shubnikov-de Haas oscillations in the magnetoresistivity. The above effects were repeatedly used in studies of the FS geometries and other electronic properties of various conventional metals \cite{1}. In the last three decades quasi-two-dimensional (Q2D) materials with metallic-type conductivity (intercalated compounds, organic metals and some other)     attracted a significant interest of the research community which resulted in extensive studies of their electronic characteristics. These materials reveal strong anisotropiy of the electrical conductivity which reflects their layered structure. Conducting layers are rather weakly coupled to each other, so the charge carriers energy only slightly depends on the momentum projection on the normal to the layers. Correspondingly, the Fermi surfaces of Q2D metals could be described as systems of weakly corrugated cylinders \cite{2,3,4}. Again, magnetic quantum oscillations were widely used for obtaining important band-structure parameters for these materials. A theory of de Haas-van Alphen and Shubnikov-de Haas oscillations in Q2D conductors is proposed in several works (see e.g. Refs. \cite{5,6,7,8,9}). 

So far, much less attention was paid  to quantum  oscillations in the elastic response of a Q2D metal to an external deformation  generated by a traveling sound wave. Such oscillations in the sound velocity and attenuation are known in conventional metals. Studies of these oscillations were started by Gurevich, Skobov and Firsov \cite{10} who first predicted the effect of giant quantum oscillations in the ultrasound attenuation and extended in several works \cite{11,12,13}. Magnetic quantum oscillations in the sound velocity and attenuation were repeatedly observed in 3D metals \cite{1}.

It has been shown both theoretically and experimentally that the specific geometry of the Fermi surfaces of Q2D metals may cause significant differences in the size, shape and phase of the de Haas-van Alphen oscillations compared to those occurring in conventional metals. Therefore, one may expect similar features in the elastic response quantum oscillations in Q2D conductors to occur. The purpose of the present work is to analyze these features. We concentrate on the case of the high frequency ultrasound waves $(\omega\tau > 1$ where $\omega $ is the wave frequency and $\tau $ is the scattering time for the charge carriers). The high frequency range is chosen for it provides opportunities for richer and more complicated structure to be revealed in the oscillating corrections to the ultrasound velocity and attenuation.

When a sound wave propagates through a metal, the crystalline lattice is periodically deformed, which brings changes to the electronic spectrum. These changes could be allowed for by introducing the deformation potential but here we omit them for brevity. Besides, the conduction electrons (charge carriers) affect the crystalline lattice by interaction with the self-consistent alternating electric field accompanying the sound wave as it travels in the metal. As a result, electron contributions appear in the expressions for elastic  constants of the metal. When a strong external magnetic field is applied, these terms include corrections describing quantum oscillations in the elastic response of the metal to the sound wave.

To analyze the magnetic quantum oscillations in the elastic response of a Q2D metal we adopt the commonly used simple approximation for the charge carriers spectrum: 
 \be
 E {\bf (p)} = \frac{{\bf p}_\perp^2}{2m_\perp} - 2t\cos \left(\frac{\pi p_z}{p_0} \right).   \label{1}
 \ee
 Here, $z $ axis is taken to be  normal to the conducting layer plane, $ {\bf p}_\perp $ is the momentum projection in the layer plane, and $m_\perp $ is the effective mass corresponding to the charge carriers motion in this plane. The parameter $ t $ in  Eq. (\ref{1}) is the  interlayer transfer integral, and $ p_0 = \pi\hbar/d $ where $ d $ is the distance between the layers. When a quantizing magnetic field $ \bf B $ is applied along the normal to the layers, the Landau energy spectrum of the charge carriers has the form:
   \be
    E_{n}^{\sigma} (p_z) = \hbar\Omega \left(n + \frac{1}{2}\right) + \frac{\sigma}{2}g\hbar \Omega_0 - 2 t\cos \left(\frac{\pi p_z }{p_0} \right).  \label{2}
 \ee
  Here, $ \Omega $ is the cyclotron frequency, $ \Omega_0 = \beta B;\ \beta $ is Bohr magneton, $ \sigma $ is the spin quantum number and $ g $ is the spin splitting coefficient $(g$-factor).

We consider a longitudinal ultrasound wave traveling along the normal to the conductivity layers (parallel to the magnetic field) with  frequency $ \omega $ and  wave vector $ {\bf q} = (0,0,q).$ An expression for the wave vector of the sound wave can be written  as follows:
 \be
 q = \frac{\omega}{s} + \Delta q   \label{3}
 \ee
 where $ s $ is the speed of sound in the absence of the external magnetic field, and  $ \Delta q $ determines the magnetic field induced corrections to the velocity shift  $ \Delta s $ and attenuation rate $ \Gamma:$ 
 \be
 \frac{\Delta q }{q} = \frac{\Delta s}{s} + \frac{i\Gamma}{2q} . \label{4}
 \ee
 Using the general equations for the magnetoacoustic response of a metal \cite{14,15} we can obtain the following expression for the correction $ \Delta q:$  
 \be
  \frac{\Delta q }{q} = -\frac{N^2 }{2\rho_m s^2} \frac{1}{\eta} Y \label{5}
 \ee 
  where $\rho_m,\ N $ are the density of matter in the lattice and the  charge carriers density, respectively, and $ \eta $ is the charge carriers DOS on the Fermi surface in the absence of the magnetic field.
  The function  $ Y $ describing quantum oscillations of the elastic response has the form:
  \begin{align}
 Y =& \frac{1}{4\pi^2\hbar\lambda^2} \frac{1}{\eta}
 \nn\\ &\times
 \sum_{n,\sigma} \int_{-p_0}^{p_0} dp_z \frac{f_n^\sigma (p_z) - f_n^\sigma (p_z - \hbar q)}{E_n^\sigma (p_z - \hbar q) - E_n^\sigma (p_z) + \hbar\omega + i\hbar/\tau }    .
   \label{6}
 \end{align}
 Here, $ f_n^\sigma (p_z) $ is the Fermi distribution function for the quasiparticles with the energies $ E_n^\sigma (p_z)$, and $ \lambda $ is the magnetic length.

In further consideration, we assume that the cyclotron quantum $ \hbar\Omega $ is small compared to the chemical potential of the charge carriers $ \mu. $ Then we employ the Poisson summation formula:
  \be
 \sum_{n=0}^\infty \varphi \left(n + \frac{1}{2} \right) = \int_0^{\infty} \varphi(x) \Big[1 + 2 \mbox{Re} \sum_{r=1}^{\infty} (-1)^r \exp (2\pi irx)  \Big].  \label{7}
 \ee
  Using this formula we may present the function $ Y$ as a sum of a monotonic term $Y_0$ and an oscillating correction $ \Delta.$ Strong anisotropy in the transport characteristics of Q2D conductors implies a pronounced difference between the quasiparticles velocities in the layer planes $ v_\perp $ and the longitudinal velocity $ v_{z}.$ Due to the smallness of the transfer parameter $ t, $ the longitudinal component of the charge carrier velocity on the Q2D Fermi surface may take on values significantly smaller than those typical for Fermi velocities in conventional metals. 

In consequence, the ratio $ s/v_{z} $ in Q2D conductors may take on greater values,  than in usual 3D metals, and we may expect an inequality $ (s/v_{z}) \sqrt{\mu/\hbar\Omega} >1 $ to be satisfied when $v_z$ takes on its maximum value  $v_0.$ Then the oscillating function $ \Delta $ may be approximated by the expression:
  \begin{align}
 \Delta = &- \frac{m_\perp}{2\pi^2 \hbar^3} \frac{1}{\eta} \sum_\sigma \sum_{r=1}^\infty (-1)^r \int dE \frac{df^\sigma (E)}{dE}
 \nn\\ &\times
 \int_{-p_0}^{p_0} dp_z \frac{v_z}{v_z - s - i/q\tau} \cos \Big[ r \frac{\lambda^2}{\hbar^2} A_\sigma (E,p_z) \Big]  \label{8}
  \end{align}
  where the longitudinal velocity $ v_z $ is given by:
  \be
 v_z = \frac{\partial E}{\partial p_z} = \frac{2\pi t}{p_0} \sin \left(\frac{\pi p_z}{p_0} \right)  \label{9}
  \ee 
  and $A_\sigma (E,p_z)$ is the cross-sectional area of the corresponding constant-energy surface. To properly estimate the integral over $ p_z $ in  Eq. (\ref{8}) we extend the integrand over the upper half of the complex plane, and we choose the integration path including the segment of the real part axis $ -p_0 \leq p_z \leq p_0 $ and the circular arc with the radius $p_0. $ The value of the integral crucially depends on the contribution from a pole at $ v_z = s + i/q\tau $ which may be  situated within the integration contour.

For a weakly warped Fermi surface $(2\pi t/p_0 < s)$ the integrand is an analytic function over the area  within the integration path. Then the value of the of the integral over $``p_z" $ in  Eq. (\ref{8}) is solely determined by the contribution from the arc. Computing the latter we obtain:
 \begin{align}
  \Delta \approx & -\frac{\alpha(ql)^2}{(1- i\omega\tau)^2} \sum_{r=1}^\infty (-1)^r D(r) \cos \left( \frac{2\pi rF}{B} \right)
 \nn\\  &\times \left[ J_0 \left(\frac{4\pi rt}{\hbar\Omega}\right) + J_2 \left(\frac{4\pi rt}{\hbar\Omega}\right) \right].  \label{10}
  \end{align}
 Here, $\ds \alpha = \frac{m_\perp p_0}{\pi^2\hbar^3}\frac{1}{\eta};\ l = v_{0} \tau $ is the mean free path of the charge carriers along the normal to the conducting layers; $J_{0,2} (x) $ are the Bessel functions;$ F = cA/2\pi\hbar e;\ A$ is the mean cross-sectional area of the Fermi surface. The damping factor $D(r)$ includes the effects of temperature, scattering and spin splitting.

\begin{figure}[t]  
\begin{center}
\includegraphics[width=7cm,height=5.5cm]{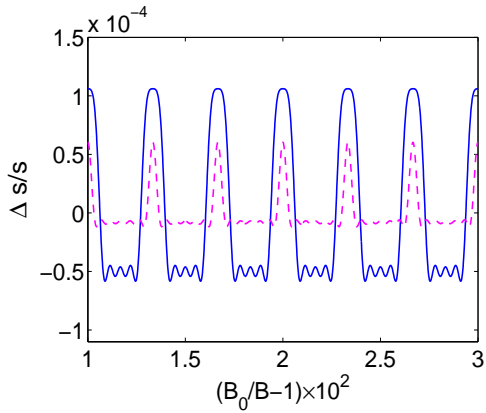}
\end{center}
\caption{(Color online) Magnetic quantum oscillations of the velocity of a longitudinal ultrasound wave propagating along the symmetry axis of a weakly corrugated Fermi surface in a Q2D metal. The plotted curves are described by Eqs. (\ref{5}) and (\ref{8}) assuming that $F/B_0 = 300,\ B_0 = 10T,\ t/\hbar\Omega(B_0) = 0.1 $ (solid line) and $ 0.04 $ (dashed line).}%
\label{rateI}%
\end{figure}

The obtained expression (\ref{10}) describes magnetic quantum oscillations in the sound velocity and attenuation originating from the DOS oscillations in the strong magnetic fields. An explicit expression for the oscillating part of $ \Delta q $ may be written out by substituting Eq. (\ref{10}) into Eq. (\ref{5}). Oscillations in the velocity of sound described by Eq. (\ref{10}) are shown in  Figure 1 assuming that the effects of temperature and scattering are moderately small. The oscillations look like a sequence of well distinguishable peaks separated by nearly flat regions. As the FS warping decreases, the peaks become sharper and  lower in magnitude. We remark that the function $ \Delta $ as given by the Eq. (\ref{10}), goes to zero when the FS becomes purely cylindrical $(t \to 0).$ This is a reasonable result for at $ t=0$ charge carriers cannot move between the conducting layers in the Q2D metal and, consequently, they cannot respond to the sound wave travelling across the latter.

Now, we consider the case of a more pronounced warping of the Fermi surface $(2\pi t/p_0 > s). $ Then the integrand in the integral over $``p_z"$  in  Eq. (\ref{8}) has the pole within the contour of integration at $ p_z = s + i/q\tau. $ Correspondingly, the oscillating function $ \Delta $ equals the sum of the residue from the pole $ \Delta_1$ and the contribution from integration over the arc $ \Delta_2. $ These terms could be approximated by the following expressions:  

  \begin{align}
 \Delta_1 \approx & - \frac{\alpha\pi i}{2} \frac{s}{v_{0}} \sum_{r=-\infty}^\infty (-1)^r D(r) 
 \nn\\ & \times  \exp\left(-2\pi ir \frac{F}{B}\right)
 \cos\left(\frac{4\pi rt}{\hbar\Omega}\frac{s^2}{v_{0}^2}\right)         \label{11}
 \end{align}
  \be
 \Delta_2 \approx \alpha \sum_{r=1}^\infty (-1)^r D(r) J_0 \left(\frac{4\pi rt}{\hbar\Omega} \right) \cos\left(2\pi r\frac{F}{B}\right). \label{12}
 \ee
 Here, the term $\Delta_2$ describes oscillations directly related to the DOS oscillations in a quantizing magnetic field. The origin of the term $ \Delta_1 $ is rather more complicated, and its occurrence needs some particular explanation. As follows from the form of the charge carriers spectrum in a strong magnetic field given by the Eq. (\ref{2}), the quasiparticles energies  corresponding to their motions in the planes perpendicular to the magnetic field are quantized. Therefore at a certain magnitude of $ \bf B$ the longitudinal part of a quasiparticle energy at the Fermi surface can take on values belonging to a set of intervals with widths of the order of the thermal energy $ kT\ (k $ is the Boltzmann's constant). 

This leads to a similar distribution of the possible values of the longitudinal velocity $ v_z = \partial E /\partial p_z $ at the Fermi surface. Within the low temperature limit $ (T\to 0)$ the intervals are reduced to points, so we obtain a discrete set of possible $ v_z $ values for every magnitude of the field $ \bf B.$ Quasiparticles on the Fermi surface may absorb phonons provided that the conservation laws for both energy and momentum are satisfied. This occurs when the longitudinal velocity $ v_z $ coincides with the speed of sound $ s. $ At certain magnitudes of the magnetic field we may find $ v_z = s $ among the allowed values of the longitudinal velocity, so the relevant quasiparticles may absorb the phonons. A slight change in $ B $ shifts all allowed values of $ v_z $ and leads to a situation where there is no charge carriers at the Fermi surface capable of phonon absorption. Varying the magnetic field magnitude we successively satisfy and destroy the ``resonance" conditions for the phonon absorption generating the giant quantum oscillations in the ultrasound attenuation. Such oscillations are described by the imaginary part of the function $ \Delta_1, $ which agrees with the well known result of Ref. \cite{8}. Sound attenuation oscillations in a Q2D conductor display a sequence of very sharp peaks separated by regions of much weaker attenuation resembling those observed in conventional 3D metals. This is shown in Fig. 2.

\begin{figure}[t]  
\begin{center}
\includegraphics[width=7cm,height=5.5cm]{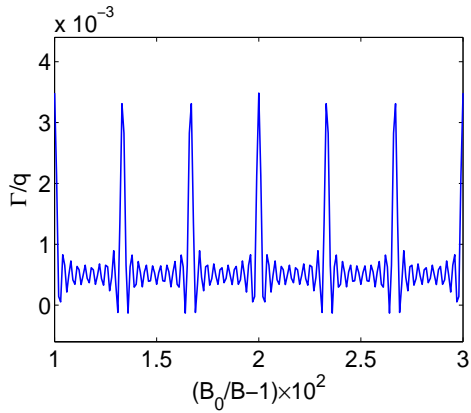}
\end{center}
\caption{(Color online) Quantum oscillations in the attenuation of the longitudinal high frequency sound propagating along the symmetry axis of a moderately corrugated Fermi surface in a Q2D metal. The curve is plotted assuming that $F/B_0 = 300,\ B_0 = 10T $ and $t/\hbar\Omega(B_0) = 4.$}%
\label{rateI}%
\end{figure}

\begin{figure}[t]  
\begin{center}
\includegraphics[width=7cm,height=5.5cm]{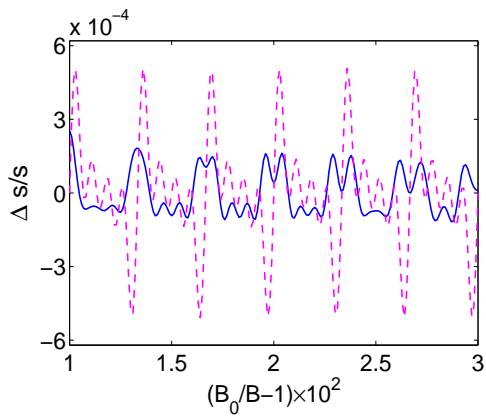}
\end{center}
\caption{(Color online) Oscillating corrections to the velocity of a longitudinal sound wave travelling along the symmetry axis of a moderately corrugated Fermi surface of a Q2D metal. The plotted curves are described by Eqs. (\ref{5}), (\ref{11}), (\ref{12}) for $ F/B_0 = 300.\ B_0 = 10T, \ t/\hbar\Omega(B_0) = 4.$ The solid line corresponds to the term proportional to $\Delta_2 $ and appears due to  quantum oscillations of the charge carriers DOS at the Fermi surface in a strong magnetic field. The dashed line displays the correction proportional to $\Delta_2 $ and originates from the ``phase stability" resonance occurring when the longitudinal velocity of some charge carriers at the FS coincides with the sound velocity. }%
\label{rateI}%
\end{figure}

Besides the imaginary part, the function $ \Delta_1 $ has  a nonzero real part accounting for an extra term in the oscillating correction to the sound velocity. This correction describes quantum oscillations controlled by the same mechanism as the giant quantum oscillations in the ultrasound attenuation. The magnetic field dependencies of the above oscillating terms in are presented in Fig. 3. We see that both shapes and phases of the the oscillations described by $ \Delta_1 $ (dashed line) and $ \Delta_2 $ (solid line) differ. A rather complicated pattern of the DOS related oscillations originates from the Fermi surface  rippling. Actually, the solid line in  Figure 3 appears as a superposition of contributions from the FS cross-sections with minimum and maximum areas. The resulting dependence of the oscillating correction to the speed of sound wave including both corrections is shown in Fig. 4.

In conclusion, in this work we theoretically analyzed magnetic quantum oscillations in the elastic response of a Q2D metal to a high frequency ultrasound wave. Within a chosen geometry a longitudinal ultrasound wave was assumed to travel along the magnetic field directed in parallel with the axis of symmetry of the Fermi surface. We showed that both sound velocity and attenuation reveal quantum oscillations provided that the FS corrugation is not too weak. The case of moderately strong Fermi surface warping $(t > s p_0)$ is especially interesting. We showed that in this case the oscillating correction to the sound velocity includes two terms of different origin and this is the main result of the present work. At low temperatures $(kT < \hbar\Omega)$ the two terms significantly differ in shape and phase. Also, their periods are slightly different, as follows from Eqs. (\ref{11}) and (\ref{12}). An extra term in the oscillating correction to the velocity of sound proportional to the function $ \Delta_1 $ appears due to some kind of ``phase stability" resonance occurring when a charge carrier moves along the magnetic field at the same velocity as the sound wave travels.
  Therefore, the relevant quasipartices are seeing the same phase of the wave all the while between consequtive collisions, and can efficiently absorb phonons. So, the above oscillations in the sound velocity appear due to the same reason as giant quantum oscillations in the sound attenuation studied in 3D metals. In whole, the present results enable us to better understand specific features of magnetic quantum oscillations in the elastic response of Q2D conductors, and they could be easily generalized to describe 3D metals.   

\begin{figure}[t]  
\begin{center}
\includegraphics[width=7cm,height=5.5cm]{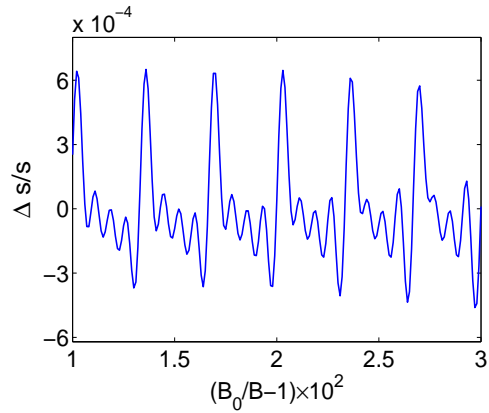}
\end{center}
\caption{(Color online) Superposition of the quantum oscillations described by Eqs. (\ref{5}), (\ref{10}), and (\ref{11}). The relevant parameters take on the same values as in the previous figure. }%
\label{rateI}%
\end{figure}

{\it Acknowledgement:}
We  thank   G.M. Zimbovsky for his help in manuscript.  NZ gladly acknowledges support from NSF-DMR-PREM 0353730 and DoD grant W911NF-06-1-0519.

\end{document}